\documentstyle[12pt]{article}

\newcommand{\be}{\begin{equation}}
\newcommand{\ee}{\end{equation}}

\newcommand{\mev}{\rm MeV}
\def\journalfont{\it}      
\def\jou#1{{\journalfont #1\ }}
\def\pl{\jou{   Phys.\ Lett.}}
\def\pr{\jou{   Phys.\ Rev.}}

\def\apj{\jou{  Astrophys. J.}}
\def\ut#1{\mathop{\vtop{\ialign{##\crcr
     $\hfil\displaystyle{#1}\hfil$\crcr\noalign
     {\kern1pt\nointerlineskip}\hbox{$\hfil\sim\hfil$}\crcr
     \noalign{\kern1pt}}}}}
\def\undersim{\ut}

\begin{document}
 \begin{titlepage}
 \parindent 0pt
 \font\big=cmbx10 scaled\magstep2
astro-ph/yymmnn\\
Uppsala-UUITP 14/1994\\
September, 1994
\vskip 1cm
\begin{center}
{\big Limits on possible magnetic fields at nucleosynthesis\\
time}
\vskip .5cm
Dario Grasso$^{\dag}$
 \vskip .3cm
 Department of Theoretical Physics, Uppsala University\\
Box 803 S-751 08 Uppsala, Sweden
\vskip0.6cm
H. R. Rubinstein
 \vskip .3cm
Department of Theoretical Physics, Uppsala University\\
Box 803 S-751 08 Uppsala, Sweden
\end{center}
\vskip .5cm
\begin{abstract}
In this paper we discuss limits on magnetic fields that could have been
present at nucleosynthesis time. We considered several effects
that could be relevant modifing light elements relic abundances.
They include: changes in reaction rates, mass shifts due to strong and
electromagnetic interactions, variation of the expansion rate of the
Universe due to both the magnetic field energy density and the increasing
of the electrons density in overcritical magnetic fields.
 We find that the latter is the main effect. It was not taken into account
in previous calculations. The allowed field intensity at the end of
nucleosynthesis ($T =1\times 10^9\;^o{\rm K}$) is
$B \le 3\times 10^{10}$ Gauss.
\end{abstract}
\vskip 1cm
{\it Accepted by Astroparticle Physisc}
\vskip 2cm
$^{\dag}$ EEC Twinning Fellow
 \end{titlepage}
\newpage
\font\big=cmbx10 scaled\magstep1
\font\vbig=cmbx10 scaled\magstep2
\vskip .2cm
\noindent
\section{Introduction}
\vskip .2cm
Amongst the many uncertainties in the Early Universe environment, the
possibility that large, constant magnetic fields, existed over macroscopic
scales, is a fascinating possibility. Since the Early Universe is believed
to be a perfect conductor, magnetic lines get thinned out by the ratio
\begin{equation}
\frac{B_1}{B_2}=\left(\frac{R_2}{R_1}\right)^2 \label{1}
\end{equation}
The presence of primordial fields over galactic scales, when extrapolated
back can give, under different assumptions, very large fields indeed.\\
A typical present day galactic field $B=10^{-6}$G, can grow, scaled as
dictated by Eq.(\ref{1}) alone to be as large as $10^{14}$ Gauss.
These extrapolations are very doubtful
since dynamo effects may have enlarged significantly present fields. \\
Nevertheless, knowing the allowed fields at a given epoch and limiting its
value at another one can give important dynamical restrictions.\\
Recently Vachaspati \cite{Vachaspati} has shown that
large magnetic fields, of the order of $B=(m_W)^2 = 10^{24} $ Gauss,
should be generated at the electroweak phase transition time.
His argument seems quite general.
It only depends on the finiteness of the horizon and embedding
electromagnetism in a larger field.
The size of a patch at electroweak time, is of order $\chi =N(m_W)^{-1}$.
Evolving the size of the patch is model
dependent. More precisely, the coefficient N in \cite{Vachaspati}
is a function of the scale.
If the scale is determined by today's size then $N=10^{13}$.
This argument predicts a field of about a Gauss at nucleosynthesis time.\\
The size of the patch at the nucleosynthesis time might allow for larger
fields, though we have no real reliable model for the field evolution. \\
In this paper we address ourselves to the allowed magnetic fields at
nucleosynthesis time, without discussing their origin. We give a detailed
analysis of the influence of the fields  on the main quantities that can
act to modify the relic elements abundance ratios: reaction rates,
masses of the participants, electron energy densities and magnetic
field energy density. We then obtain, given the present errors in these
relative abundances, the upper limits the fields can take in regions large
compared to
the reaction scale but possibly much smaller than the horizon at that time.
In the first section we discuss briefly the impact of the magnetic field on
these elements of the calculation. In the next we describe the standard
nucleosynthesis calculations in this light. In the final section we discuss
the constraints and their origin and compare with existing calculations.
\vskip 2 cm

\section{Weak reaction rates in the presence of magnetic fields}

The main weak processes which act to determine the $n/p$ ratio during the
the primordial nucleosynthesis are
$$ n \,\, + \,\, e^+ \,\, \rightleftharpoons \,\, p\,\, + \,\,\overline\nu ,
\eqno(a)$$
$$ n \,\, + \,\, \nu \,\, \rightleftharpoons \,\, p\,\, + \,\,e^- ,
\eqno (b)$$
$$ n \,\, \rightleftharpoons \,\, p\,\, + \,\,e^- \,\, + \,\,\overline\nu
.\eqno (c)$$
The rate for two body scattering reactions in a medium may be
written in the form
\be
\Gamma(12 \rightarrow 34) =
\left(\prod_i \frac{\int d^3 {\bf p}_i}{(2 \pi)^3 2 E_i} \right)
  (2 \pi)^4 \delta^4(\hbox{$\sum_i$} p_i) |{\cal{M}}|^2\,
    f_1 f_2 (1 - f_3) (1 - f_4),
\ee
where $p_i$ is the four momentum, $E_i$
is the energy and $f_i$ is the number density of each particle species.
All processes in Eq. (a,b and c) have the same amplitude
\be
{\cal{M}} = \frac{G_F}{\sqrt{2}}\cos\theta_C\;{\bar u}_p\gamma_\alpha (1 -
\alpha \gamma_5) u_n{\bar u_e}\gamma_\alpha (1 - \gamma_5) u_{\nu}
\ee
where $\alpha = g_A/g_V \simeq = -1.262$.
Without any external magnetic field the total rate of the processes
that convert neutrons to protons is
\be
\Gamma_{n\to p}(B=0) = {1\over \tau} \int_1^\infty d\epsilon\,{\epsilon
\sqrt{\epsilon^2 - 1} \over 1+e^{{m_e\epsilon\over T}+\phi_e} }
\left[{(q+\epsilon)^2
e^{(\epsilon + q )m_e\over T_\nu}\over 1+e^{(\epsilon + q) m_e\over T_\nu}}
+ {(\epsilon-q)^2 e^{{\epsilon m_e\over T}+\phi_e } \over
1+e^{(\epsilon-q)m_e\over T_\nu}}\right] \label{rate0}
\ee
where ${1\over \tau} \equiv {G^2 (1 + 3 \alpha^2) m_e^5 \over 2 \pi^3}$
and $q$, $\epsilon$ and $\phi_e$ are respectively the neutron-proton mass
difference, the electron energy and
the electron chemical potential all expressed, in units of $m_e$.
We assume the neutrinos chemical potential to be vanishing.

 The total rate for the $p \to n$ processes can be obtained changing the
sign of $q$ in Eq.(\ref{rate0}).

An external magnetic field take us to modify Eq. (\ref{rate0}) due to
the following effects.

{\indent{a)}
The dispersion relation of charged particles propagating trough a
magnetic field is modified with respect to the free-field case.
In fact, their 4-momentum is in this case
$p  = p(B=0) + q A$ ,where q is the charge of the particle and the vector
potential ${\bf A(r)}$ is related to
the field by ${\bf A(r)}={1\over 2}{\bf r\times B}$.
Assuming  ${\bf B}$ along the ${\bf {z}}$ axis, the expressions for
energies of electrons, protons and neutrons are respectively
\be
E_e = \left[p_{e\,,z}^2 + eB(2n+1+s) + {m_e}^2 \right]^{1\over 2} + \kappa
 \; ,\label{disp}
\ee
\be
E_p = \left[p_{p\,,z}^2 + eB(2n+1-s) + {m_p}^2 \right]^{1\over 2} -
{e\over {2m_p}}\left({g_p\over 2}-1\right)B\; ,
\ee
\be
E_n = \left[{\bf p}_{n}^2 + {m_n}^2 \right ]^{1\over 2} +
{e\over 2m_n}{g_n\over2}B\; .
\ee
In the above, $n$ denotes the Landau level, $s=\pm 1$ indicates
whether the spin is along or opposed to the field direction, and
$g_p = 5.58$ and $g_n = -3.82$ are the Land\'e g-factors.
The QED correction to the electron energy, $\kappa$, has been first computed
by Schwinger \cite{Schwinger}. For magnetic fields larger
than $\sim  10^{13}\;{\rm G}$ this correction is
\be
\kappa = {\alpha\over 2\pi}\,\ln\left({2eB\over m_e^2}\right)^2\ .
\ee
For smaller field intensity $\kappa$ has negligible
effects on our calculations and we disregarded it.
 The effects of the field on the QCD ground state have been parametrized
via a field dependent nucleon mass \cite{Bander}, as we are going to
discuss below.

Neither the neutron or the neutrino have quantized levels, though the
neutron has an electromagnetic interaction energy. The neutrino is totally
inert vis a vis electromagnetism.

{\indent{b)}
The number of available states for a particle obeying Eq. (4)
becomes, for every value of $n$ and $s$, is\cite{Landau}
\be
{V\, eB\over (2\pi)^2}\,dp_z \label{phase}
\ee
This changes the phase space of the processes we are interested to consider.

{\indent{c)} Since the occupation number and the energy of states with
opposite spin projections is not the
same in a magnetic field, the spin sum of the square amplitude needs to be
weighted by the appropriate spin dependent Fermi distributions.

Nucleosynthesis take place in a range of temperatures $0.1 < T <
10\;\mev$, hence nucleons are nonrelativistic. Therefore nucleon
distribution functions are given by
\be
f_N(s=\pm 1) = (1 + e^{\mp {\mu_N B\over T}})^{-1}
\ee
where
\be
\mu_p = {e\over 2m_p}\,{g_p\over 2}\qquad
\mu_n = {e\over 2m_n}\,{g_n\over 2} \;.
\ee
 Since during the nucleosynthesis $m_N \gg T$, and momenta are
also small compared to nucleon mass, $f_N$ can be
safely approximated by $1/2$. This is not the case for electrons.
In this case we have
\be
f_e(s) = \left(1 + e^{E_e(s)/T}\right)^{-1}
\ee
where the relativistic expression for the electron energy Eq.(\ref{disp})
is used.

As a consequence, the integral for the leptonic momentum space in
neutron $\beta$ decay is modified to
\be
{1\over 2\pi} \sum_{n=0}^{N_c}\int_{-\infty}^\infty
{d^3{\bf p}_\nu \over 2E_\nu}\,
\int_{-p_{e,z}(n)}^{p_{e,z}(n)}{dp_{e,z}\over 2E_e}eB\,|{\cal M}|^2
(1 - f_\nu)(1 - f_e)
\ee
where $N_c$ is largest integer $n$ such that
$p_{e,z}(n)^2 = Q^2 - m_e^2 - 2n eB$ is positive and $Q^2 \equiv
m_n^2 - m_p^2$

{\indent{d)} The nucleon masses are affected by very strong magnetic fields.
The change in effective phase space is \cite{Bander}
\be
\Delta= 0.12\mu_N B - M_n+ M_p + f(B) \; .
\ee
The function $f(B)$ gives the rate of mass change due to colour forces being
affected by the field. For nucleons \cite{Bander} the main change is the
chiral condensate growth, which because of the different quark content
of protons and neutrons makes the proton mass grow faster\cite{Bander}.
Though the sign is certain, vacuum pairs
of zero helicity get more bound in the presence of a B field, the size of
the effect is model dependent. We have calculated its influence using the
weakest and strongest reasonable field dependence and we find the effect
always small for fields below $10^{18} $ Gauss.

Having established that hadronic mass changes will not affect
nucleosynthesis we drop these effects from the equations altogether.

Taking into account the remaining effects we computed the total rate for the
weak processes converting neutrons to protons in an external magnetic field.
The result is
$$\Gamma_{n\,\to\,p}(B) =
 {\gamma\over \tau}\sum_{n=0}^{\infty} (2 - \delta_{n0})
 \times \int_{\sqrt{1+2(n+1)\gamma} + \kappa}^\infty \;d\epsilon
{(\epsilon - \kappa)\over \sqrt{(\epsilon -\kappa)^2-1-2(n+1)\gamma}}
$$
\be
 \times {1\over 1+e^{{m_e\epsilon\over T} + \phi_e}} \left[
{(\epsilon+q)^2 e^{m_e(\epsilon+q)\over T_\nu} \over 1+e^{{m_e(\epsilon + q)
\over T_\nu} + \phi_e}}
 + {(\epsilon-q)^2 e^{{m_e\epsilon \over T} + \phi_e}\over
1+e^{m_e(\epsilon-q)\over T_\nu}}\right] \label{rate}
\ee
where $\gamma \equiv B/B_c$ and $B_c = m_e^2/e = 4.4\times 10^{13}\,G$ is
usually defined to be the critical magnetic field.

Equation (\ref{rate}) coincides with the result of Matese and O'Connell
\cite{Matese} and Cheng et al. \cite{Schramm1} in the limit in which
the QED correction $\kappa$ goes to zero. Although the quantitative effects
of this term on the nucleosynthesis predictions are subdominant, we
stress that disregarding it when the field is overcritical it leads
to an unstable electron ground state, thus to unphysical results.

The main effect of the magnetic field is due to the modification of the
electron phase space.
Eq.(\ref{rate}) is correct in the weak field limit, when $\gamma \ll 1$ and
$\Gamma_{n\,\to\,p}(B)$ reduces to Eq.(\ref{rate0}) in the $B=0$ limit.

In Fig.(1) we present the dependence of $\Gamma_{n\,\to\,p}$
as a function of temperature for some values of $\gamma$.
%

For large values of $\gamma$ and fixed temperature, the total rate grows
like $\gamma$. Increasing the temperature the relevant contribution to the
integrals in Eq.({\ref{rate}) comes from the high energy part of the
electron spectrum. Since the limit $\epsilon \rightarrow \infty$ is
equivalent to the limit $\gamma \rightarrow 0$ in Eq.(\ref{rate}) this
explains why the ratio $\Gamma_{n\,\to\,p}(B)/\Gamma_{n\,\to\,p}(0)$ goes to
one when $T \gg m_e$.
Although the global rate of the inverse process $\Gamma_{p\,\to\,n}$ also
increases with $B$, it remains suppressed by a factor $\exp(-Q(B)/T)$ with
respect to $\Gamma_{n\,\to\,p}$. Thus the effect of a strong magnetic field
would be to {\it reduce} the final number of neutrons in the Universe,
i.e. the relic $^4$He abundance, if only the correction to the weak rates is
taken into account.

\section{The effects of $\bf B$ on the expansion rate}

Owing to exponential dependence of the $(n/p)$ equilibrium ratio on the
temperature, the relic relative abundances of light elements depends
crucially on the freeze-out $T_F$ temperature of the weak processes
that keep protons and neutrons in chemical equilibrium \cite{KolbT}.
This temperature is essentially determined by the condition
\be
\Gamma_{n\rightleftharpoons p}(T_F) =  H(T_F)
\ee
where $H$ is the expansion rate of the Universe.

It is evident that besides the rate of the weak processes we need to pay
attention to the effects of the magnetic field on $H$.
 If no cosmological constant is present, the expansion rate is determined
by the Einstein equation
\be
H^2(T) = {8\pi G_N\over 3} \rho(T)\ .
\ee
where $\rho(T)$ is the total energy density of the Universe.
In the case that no magnetic field is present $\rho(T)$ is given by the
sum of the energy density of all the particle species in thermal
equilibrium with the primordial plasma
\be
\rho(T) = \rho_\gamma(T) + \rho_e(T) + \rho_\nu(T)
+ \rho_b(T) \label{rho}
\ee
where the subscripts $\gamma, \,\, e,\,\, \nu,\,\,b$
stand, respectively, for photons, electrons, the three species of neutrinos
and baryons, including their respective anti-particles.
In our case, since the magnetic field has energy density
$\rho_B(T) = B(T)^2/8\pi$ this term also needs to be added
to Eq.(\ref{rho}).
Since we have magnetic flux conservation in the plasma
$$\quad B\propto R^{-2} \propto T^2$$
the energy density of the magnetic
field has the same temperature dependence as the energy density of the
radiation.

This new contribution to $\rho(T)$ will dominate over the other terms
in Eq.(\ref{rho}) if
\be
B(T = 10^{11}\,^o{\rm K}) \undersim{>} 10^{16}\,{\rm G}. \label{critical}
\ee
 We assumed the pressure associated to the random magnetic field to be zero
in average. Although a novanishing mean pressure is also possible for
random magnetic fields \cite{Enqvist}, our final conclusions are not
affected also taking this pressure into account.

The presence of a nonvanishing $\rho_B$ is not the only effect that
modifies the expansion rate of the Universe.
The energy density of charged particles in the primordial
plasma is also affected. In the previous section we have showed how
the electron dispersion relation and
the electron phase-space are modified by the magnetic field.
Using Eqs.(\ref{disp}) and (\ref{phase}) we get the electron
energy density in function of $\gamma$
$$
\rho_e({\rm T}) = {eB\over 2\pi^2}\sum_{n = 0,\, s}^{\infty}\int dp_z
E_e(s) f_e = $$
\be
{\gamma\over 2\pi^2}\sum_{n = 0}^{\infty} (2 - \delta_{n0})
\int_{\sqrt{1+2(n+1)\gamma} + \kappa}^\infty \;d\epsilon\;\epsilon
{(\epsilon - \kappa)\over \sqrt{(\epsilon -\kappa)^2-1-2(n+1)\gamma}}
{1\over 1+e^{{m_e\epsilon\over T} + \phi_e}} \ . \label{rhoe}
\ee
The reader can see from Fig.(2) that the effect of an overcritical magnetic
given by Eq.(\ref{critical})
is to increase the electron energy density roughly linearly with the
field intensity. The same of Eq.(\ref{rhoe}) is valid for positrons once the
sign of the chemical potential $\phi_e$ has been changed.
%
%
%
In analogy with what happen for the reaction rates, this effect becomes
less important at high temperatures if B is left fixed. However it is a
very relevant effect when ${\rm T} \undersim{<} 1\;\mev$.
Although the field intensity, hence the correction to $\rho_e$, decreases
like ${\rm T}^2$, we are going to show that this is the main effect
on the primordial nucleosynthesis predictions.

%
%

\section{Results and conclusions}

In the previous sections we have shown that the existence of large
magnetic fields during the primordial nucleosynthesis affects the
final light elements relative abundances via two main effects:\\
{\indent{a)}} the increasing of the weak reaction rates and\\
{\indent{b)}} the increasing of the
expansion rate of the Universe. \\ These are competing effects.
In fact, whereas the former tend to reduce the $(n/p)$ freeze-out
temperature, hence the final abundance of $^4{\rm He}$, the
latter acts in the opposite direction.

We modified the standard nucleosynthesis code \cite{Wagoner}
to take into account all the relevant effects, as well as other effects
that eventually we neglected as irrelevant.\\
Since our aim is to get an upper limit to the magnetic
field intensity, we adjusted the value of the baryon photon
ratio $\eta$ in order to get the minimal $^4{\rm He}$ relic abundance
prediction compatible with observations \cite{Walker} in the free-field
case. In Tab.I we present our predictions for  some light elements
relic abundances. As it is evident at a glance, limits on magnetic fields
are totally controlled by the $^4{\rm He}$ abundance.

Other elements reach forbidden values only
at very high fields, in that case the effect of mass changes
due to colour forces will also be important.


\begin{table}
Table I. Predictions of light element abundances at the end of the
primordial nucleosynthesis are given for several values
of the magnetic field intensity given at the temperature
$T =  10^{11}\,^o{\rm K}$.
\begin{center}
\begin{tabular}{c|c|c|c}
$B(T = 10^{11}\,^o{\rm K})$ & $^4$He
& $($D$+^3$He$)/$H & $^7$Li/H \\
\hline\hline
$0$ & $0.236$ & $1.14\times 10^{-4}$ & $1.11\times 10^{-10}$\cr
$1\times 10^{12}$ & $0.236$ & $1.14\times 10^{-4}$ & $1.11\times 10^{-10}$\cr
$5\times 10^{12}$ & $0.236$ & $1.14\times 10^{-4}$ & $1.11\times 10^{-10}$\cr
$1\times 10^{13}$ & $0.237$ & $1.13\times 10^{-4}$ & $1.11\times 10^{-10}$\cr
$5\times 10^{13}$ & $0.240$ & $1.08\times 10^{-4}$ & $1.14\times 10^{-10}$\cr
$1\times 10^{14}$ & $0.242$ & $1.05\times 10^{-4}$ & $1.15\times 10^{-10}$\cr
$5\times 10^{14}$ & $0.247$ & $9.99\times 10^{-5}$ & $1.20\times 10^{-10}$\cr
$1\times 10^{15}$ & $0.250$ & $9.71\times 10^{-5}$ & $1.23\times 10^{-10}$\cr
$5\times 10^{15}$ & $0.257$ & $9.15\times 10^{-5}$ & $1.32\times 10^{-10}$\cr
$1\times 10^{16}$ & $0.348$ & $8.92\times 10^{-5}$ & $1.35\times 10^{-10}$\cr
\end{tabular}
\end{center}
\end{table}

The increase of the $^4{\rm He}$ relic abundance with the field intensity
reveals that the effect of B on the expansion rate is the most relevant.
Regarding this point we agree with the qualitative conclusion of
Matese and O'Connell \cite{Matese} and disagree with the opposite
conclusion of Cheng et al. \cite{Schramm2}. Mainly, we do not understand
how they can reconcile the claim that the dominant
effects of the magnetic field are those arising from modification of the
reaction rates, with the growing of the relic $^4{\rm He}$ that they
get increasing B.

Furthermore, in both refs. \cite{Matese} and \cite{Schramm2} the
effect of the magnetic field on the electron and positron energy density
was not considered. Leaving only this effect on in our code,
we checked that this is indeed the most relevant.

We showed that this is indeed the main effect as long
as the field intensity at the beginning of the nucleosynthesis is smaller
than $10^{16}\;{\rm G}$.

Since the observational upper bound for the $^4{\rm He}$ relic abundance
is $Y_p \le 0.245$ we conclude that the average intensity of a random
magnetic field at the temperature of ${\rm T }= 1\times 10^{11}\;^o{\rm K}$
(beginning of nucleosynthesis) must
be less than $3\times 10^{14}\;\rm{G}$ or, equivalently,
$B(T=10^9\,^o{\rm K}) < 3\times 10^{10}\;\rm{G}$ (end of nucleosynthesis).
Vachaspati \cite{Vachaspati} predicts a magnetic field strength of $\sim
10^{11}\;{\rm G}$, on the smallest coherence region of the field ($N = 1$),
at the end of nucleosynthesis. This extreme assumption ($N = 1$) is ruled
out by our limits.

Assuming the field continue to rescale according to Eq.(1)
(perhaps not a reasonable assumption), our results
imply that the intergalactic field is less than $\sim 3\times 10^{-7}
\,{\rm G}$ at present.

The other light elements relic abundances are less affected than
$^4{\rm He}$ by the magnetic field and we do not use them to get constraints.
However, it is interesting to observe the behaviour of
Deuterium and $^3{\rm He}$ abundances versus the initial magnetic field.
Although they increase with the field at the beginning of nucleosynthesis
their relic abundances follow the opposite behaviour.
This can be understood since the rates of the processes converting Deuterium
and $^3$He to $^4$He are proportional to the initial abundances $Y_D$ or
$Y_{^3{\rm He}}$ \cite{KolbT}. The greater are the rates smaller are the
freeze-out temperatures for these light elements. Thus we expect
smaller D and $^3$He relic abundances even if the relic $^4$He increases.

\section*{Acknowledgments}

 One of the authors (D.G.) is grateful to D. Fargion for pleasent and
useful discussions.

This work was supported in part by the Swedish Research Council and
a Twinning EEC contract.

\newpage
\parindent 0pt
{\vbig Figure Caption}
\vskip .7cm
\begin{itemize}
\item [Fig.\thinspace 1.]  { The neutron-depletion rate $\Gamma_{n \to p}$,
normalized to the free-field rate, is plotted as a function of the temperature
for some constant values of $\gamma$.}

\item [Fig.\thinspace 2.] {Electron energy densities are displyed
as a functions of $\gamma$. The lower curve corresponds to the temperature
value $T = 0.5\; \mev$; the middle one to $T = 1\; \mev$ and the upper one to
$T = 5\; \mev$. }

\item [Fig.\thinspace 3.]  {Electron and magnetic field energy densities
temperature dependence are compared for two different values of the initial
magnetic field.}

\end{itemize}

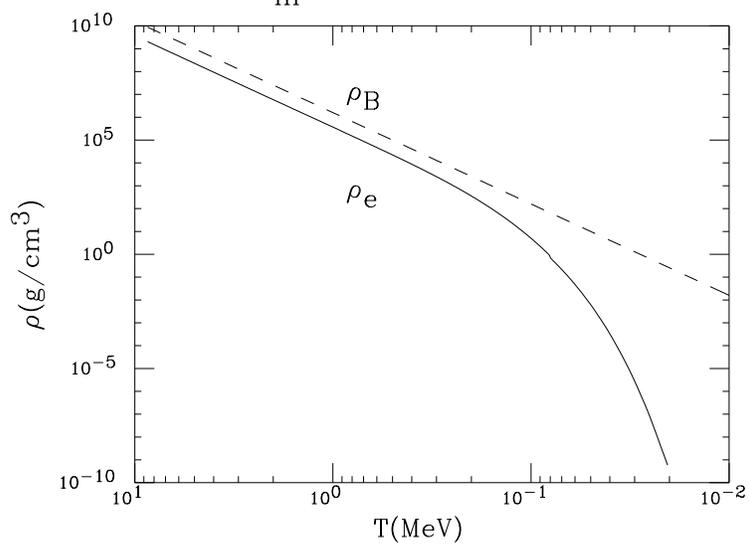
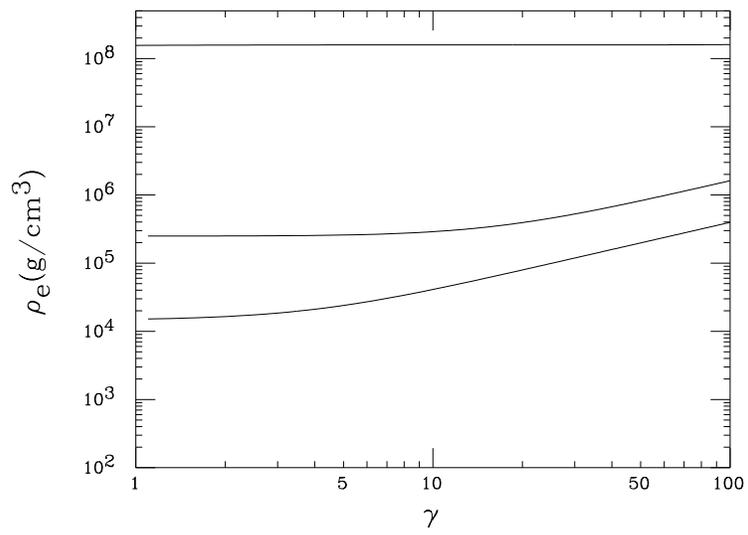
\end{document}